\documentclass{article}

\usepackage{arxiv}
\usepackage{hyperref}
\hypersetup{
  colorlinks=true,
  linkcolor=blue,      
  urlcolor=blue,
  citecolor=blue
}
\usepackage[numbers]{natbib}
\usepackage{amsmath,amssymb,amsfonts}
\usepackage{algorithmic}
\usepackage{graphicx}
\usepackage{textcomp}
\usepackage{xcolor}
\usepackage{caption}
\usepackage{url}
\usepackage{siunitx} %for table decimals number
\usepackage{braket}
\usepackage{makecell}
\usepackage{csquotes}
\usepackage{enumitem}

\title{Hybrid Data Management Architecture for Present Quantum Computing}

\author{
 Markus Zajac \\
  Databases and Information Systems\\
  FernUniversität in Hagen\\
  58097 Hagen, Germany \\
  \texttt{markus.zajac@fernuni-hagen.de} \\
  \And
 Uta St{\"o}rl \\
  Databases and Information Systems\\
  FernUniversität in Hagen\\
  58097 Hagen, Germany \\
  \texttt{uta.stoerl@fernuni-hagen.de} \\
}

\begin{document}
\maketitle

\begin{abstract}
Quantum computers promise polynomial or exponential speed-up in solving certain problems compared to classical computers. However, in practical use, there are currently a number of fundamental technical challenges. One of them concerns the loading of data into quantum computers, since they cannot access common databases. In this vision paper, we develop a hybrid data management architecture in which databases can serve as data sources for quantum algorithms. To test the architecture, we perform experiments in which we assign data points stored in a database to clusters. For cluster assignment, a quantum algorithm processes this data by determining the distances between data points and cluster centroids.
\end{abstract}

% keywords can be removed

\section{Introduction from the Database Perspective}\label{sec:intro}
Quantum computers have the potential to perform certain calculations much faster than classical computers. They can be used in various application areas, such as optimization, machine learning or search algorithms, to name just a few examples~\cite{DBLP:journals/pvldb/CalikyilmazGGWPSAPG23,Houssein22,Hassija20}. Depending on the problem, a polynomial or exponential acceleration can be assumed compared to a classical computer~\cite{Schuld.2021.3}. It is foremost a mathematical superiority. However, for practical use, certain hurdles must be addressed.

First of all, the current generation of quantum computers (NISQ) can be termed as noisy and limited in scalability~\cite{Leymann.2020,Weigold.2021c}. This means that the number of qubits is limited, and the qubits themselves are error-prone. Next difficulty concerns loading data into the quantum computer. From the perspective of a database, data is managed in various database models that need to be processed to solve a particular problem or query. In principle, it may be interesting to perform this processing on a quantum computer for certain problems or queries. We introduce some computationally intensive queries later on. Quantum computers, however, cannot access databases directly~\cite{Weigold.2021}. Moreover, today's quantum algorithms assume its input data is already in the desired form~\cite{Kieferova.2022}. The data to be processed has therefore first to be encoded in a suitable way in order to be used on a quantum computer at all. The encoding must reflect the structure of the data, for example, when data is organized hierarchically (in trees) or semi-structured (in documents). Efficient encoding of data is a challenge~\cite{Kieferova.2022,Herbert.2022} as well as a future research direction~\cite{Houssein22}. Last but not least, workflows and interfaces for data exchange between classical systems (such as applications but also databases) and quantum computers are an issue. An important design goal of such hybrid systems should be the reduction and manageability of complexity. Our contributions are therefore the following:

\begin{enumerate}
    \item In this vision paper, we describe a hybrid system that enables the exchange of data between applications and databases and quantum computers and call it Hybrid Data Management Architecture (HDMA). The architecture acts as a framework for researching and piloting appropriate encoding methods for data in different data structures and models. It is also a framework for designing future data-centric applications for quantum computing.
    \item We validate an early prototype of this architecture on an example.
\end{enumerate}

The remainder of the paper is structured as follows: In Section~\ref{sec:work} we discuss related work. In the following, we describe the HDMA (Section~\ref{sec:architecture}) and in Section~\ref{sec:trial} its proof of concept. In doing so, we use a quantum distance estimation algorithm. Section~\ref{sec:summary} provides a summary and outlines next steps.
\section{Related Work}\label{sec:work}

We review work on computationally intensive database queries, quantum technologies in the database environment, data encoding, and data exchange.

\smallskip
%\noindent
\textbf{Computationally intensive queries.} We first addressed the question of whether data models and corresponding queries exist that are of interest for quantum computing. We first look at queries that are computationally intensive and therefore represent interesting objects of investigation to determine whether quantum computers can generally enable their acceleration. As an example, data organization in labeled trees (like XML trees) can be given here. David~\cite{David.08} describes an NP-Complete problem involving data tree patterns as a query language for such trees. Gottlob et al.~\cite{Gottlob.03} discuss the complexity of \textit{XPath} queries (used for node selection) on such trees. These queries exhibit a polynomial runtime, and for some the polynomial degree can be~$4$ or~$5$. Another NP-hard problem is the keyword search in (graph) databases, which is related to the so-called \textit{Group Steiner Tree} problem~\cite{10.14778/3611540.3611577}. This non-exhaustive list of examples shows that interesting query candidates exist for quantum computing.

\smallskip
%\noindent
\textbf{Quantum technologies in the database environment.} In~\cite{DBLP:journals/pvldb/CalikyilmazGGWPSAPG23} quantum algorithms (such as \textit{Grover}) and heuristics (such as \textit{Variational Quantum Algorithms} or \textit{Quantum Annealing}) for optimizing queries and transaction plans are presented. The development of hybrid algorithms for database problems is mentioned as a future research direction. Yuan et al.~\cite{DBLP:conf/vldb/YuanLCWYYL023} also conduct a literature review on quantum computing for database problems (such as database search, database manipulation, or query optimization). In addition, the paper outlines a vision of a quantum-based multi-modal database. Jóczik and Kiss~\cite{Joczik.2020} describe some possible applications for database systems based on \textit{Grover's} algorithm. In~\cite{PhysRevA.108.032610}, the idea of a so-called data center with \textit{Quantum Random Access Memory} (QRAM) is presented. Classical or quantum information should be able to be uploaded and downloaded to the center. This uploaded information can be processed, e.g. by suitable quantum algorithms. However, the authors assume that QRAM has been constructed in a fault tolerant manner and has been error corrected. Today, however, noise resilience and scalability represent major unsolved hurdles~\cite{DBLP:journals/corr/abs-2305-01178}. An implementation of a sufficiently large memory therefore seems to be impossible today~\cite{Matteo.2020}.

In summary, the papers present visionary ideas, future research directions, and reviews of the literature. Our contribution is to develop not only concepts for the use of quantum technologies in the database environment, but also a hybrid architecture that can be implemented today to successively realize these concepts. The architecture is a framework for exploring and testing appropriate encoding methods for data in various data structures and models, and implementing certain queries using quantum computing. To the best of our knowledge, there is no other work that addresses a framework for realizing the above database queries using database systems or classical technologies and quantum technologies.

\smallskip
%\noindent
\textbf{Data encoding.} Encoding represents the transformed classical data by means of qubits~\cite{Kieferova.2022}. Only then can the data be processed on a quantum computer. In~\cite{Weigold.2021,Weigold.2021b,Weigold.2022}, various encoding procedures are explained, which are understood as patterns. These patterns are reusable self-contained building blocks that can be reused in the construction of quantum algorithms. 

As mentioned earlier, data can be organized in different data models and structural information must be preserved when encoding. A first idea of how to encode data organized in labeled trees was presented in~\cite{DBLP:conf/vldb/Zajac23}. We are not aware of any other work on encoding specific database data models. In general, the encryption methods we are developing are also intended to serve as a blueprint for future improved quantum hardware. For these, roadmaps are given, accompanied by appropriate research~\cite{Riel21}.

\smallskip
%\noindent
\textbf{Data exchange.} Another aspect concerns the exchange of data between classical components and quantum computers. Data wrangling, pre-processing, job management, post-processing, and feeding back the results to applications or a databases are individual processes here. The work of Weder et al.~\cite{Weder.2021} deals with orchestration using \textit{BPMN} workflows. The described approach with many steps and modeling aspects seems complex to us. It can be generally stated that more complex systems (such as so-called \textit{data mesh architectures}) tend towards complexity and cause considerable additional expenses for their operation~\cite{Kraska.23}.

To reduce complexity, we propose a decentralized approach with a lightweight, asynchronous communication mechanism by adopting the microservice paradigm. Microservices are typically used to manage the increase in complexity of systems~\cite{DBLP:books/sp/17/DragoniGLMMMS17}.
\section{Architectural Description}\label{sec:architecture}

\begin{figure}[ht]
 \centering
 \includegraphics[scale=0.80]{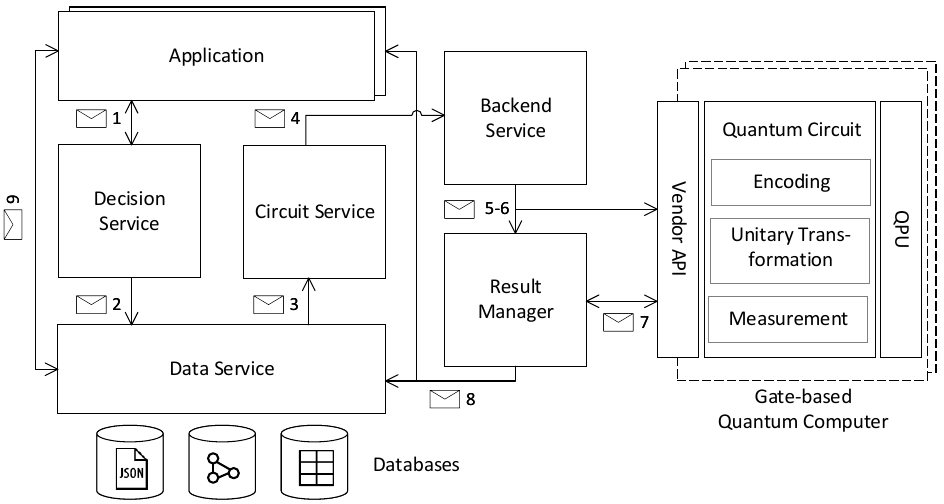}
 \caption{Hybrid Data Management Architecture [Source: Own representation]}
 \label{fig:arch}
\end{figure}

This Section introduces the Hybrid Data Management Architecture (HDMA). We derive these from the introductory Section~\ref{sec:intro} and the considered work of Section~\ref{sec:work}. Figure~\ref{fig:arch} shows the schematic structure of the architecture. The classic components include a series of loosely coupled services that follow the microservice paradigm. These are the following services: \textit{Decision Service}, \textit{Circuit Service}, \textit{Data Service}, \textit{Backend Service} and \textit{Result Manager}. Some services communicate with a \textit{Gate-based Quantum Computer} that processes data to solve a task. A \textit{Gate-based Quantum Computer} is characterised by its ability to be used for a wide range of problems. This type of quantum computer is based on quantum gates for encoding and processing data, which form a \textit{Quantum Circuit}. Today, quantum computers are usually provided as cloud services. Instead of quantum computer, a simulator can also be used.

We will first provide a brief overview of the services before focusing on communication. As motivation, we imagine an \textit{Application} that would initiate a computationally intensive query or intends to solve another computationally intensive problem. The decision as to whether a problem is calculated using a \textit{Gate-based Quantum Computer} or completely classically is made by the \textit{Decision Service}. The data required to solve a problem, as well as the results of the calculations, are stored in \textit{Databases} and managed by the \textit{Data Service}. The \textit{Circuit Service} generates \textit{Quantum Circuits} for the \textit{Gate-based Quantum Computer} that contain the data and processing instructions. The transmission of \textit{Quantum Circuits} to \textit{Gate-based Quantum Computer} is the responsibility of the \textit{Backend Service}. The \textit{Result Manager} determines whether previously transmitted \textit{Quantum Circuits} have been executed and notifies the \textit{Application} and / or \textit{Data Service} if a result is available. Based on the services and the \textit{Gate-based Quantum Computer}, we define the following reference procedure in which messages are exchanged asynchronously between the services.

\begin{description}[style=unboxed,leftmargin=0.25cm]
    \item[Message 1.] An \textit{Application} notifies the \textit{Decision Service} of the problem to be solved. This first verifies whether a problem should be calculated using a \textit{Gate-based Quantum Computer} or not and notifies the \textit{Application} of the decision. We assume in the following that a \textit{Gate-based Quantum Computer} is to be used. Otherwise, the problem is solved classically and the \textit{Application} communicates directly with the \textit{Data Service} for the purpose of data exchange (\textbf{Message~9}).
    
    \item[Message 2.] The \textit{Data Service} is notified in order to extract the data required to solve the problem.
    
    \item[Message 3.] After all data is available, the \textit{Circuit Service} is notified to generate a corresponding \textit{Quantum Circuit}. If appropriate, several circuits are created.

    \item[Message 4.] Once the circuits have been successfully created, the \textit{Backend Service} is triggered to transfer them to a \textit{Gate-based Quantum Computer}.
    
    \item[Messages 5-6.] The \textit{Backend Service} then transmits the circuits to a \textit{Gate-based Quantum Computer} using a quantum cloud service and initiates their execution (\textbf{Message~5}). At the same time, the \textit{Result Manager} is notified, which monitors the status of the execution (\textbf{Message~6}).
    
    \item[Messages 7-8.] Once the \textit{Result Manager} determines that the execution of a circuit has been completed, it retrieves the result. The status and result are retrieved via a quantum cloud service (\textbf{Message~7}). At the same time, the \textit{Application} and / or \textit{Data Service} can be notified (\textbf{Message~8}). The result can be sent directly to the \textit{Application}, stored in a \textit{Database} or both. Usually, the result obtained must be post-processed. In particular, other data records stored in \textit{Databases} can be used for this purpose. This step is performed by the \textit{Application}.  
\end{description}

Next, we take a brief look at the structure of a \textit{Quantum Circuit}. A \textit{Quantum Circuit} can be roughly divided into the areas of \textit{Encoding}, \textit{Unitary Transformation} and \textit{Measurement}~\cite{Schuld.2021.4, Weigold.2021}. The \textit{Encoding} block is responsible for encoding, which means that data and, if necessary, parameters are loaded and encoded in a quantum state. This quantum state forms the starting point for the actual quantum algorithm in the \textit{Unitary Transformation} block, which can manipulate the initial state. The execution of an algorithm ends with the \textit{Measurement} of the final quantum state, which represents the result.

\smallskip
Finally, we address two aspects that are to be supported by the HDMA.

\smallskip
\noindent
\textbf{Circuit generation.} In some cases, the circuits are generated on demand, in others automatically, e.g. when data changes. The first method can be useful if, for example, different small portions of historical data or predefined problem instances are to be processed. In the second case, data changes (e.g. new or updated data records) must be detected automatically and the circuit generation must then be triggered automatically at certain times. In this scenario, a quantum algorithm works with data set at a given time or the most recent time. 

\smallskip
\noindent
\textbf{Data restrictions / data economy.} Considering data restrictions in the architecture is central to quantum computing. More data requires more qubits and longer runtimes of state preparation routines. We propose to store data constraints in profiles to define only a reasonable minimum amount of data to be processed on a quantum computer. For a given use case, different profiles can be used, e.g. to manage different data value ranges or different graph partitions. In the latter case, different circuits could be generated for the different node sets (each partition forms the input for a calculation task). Some use cases (which especially process historical data) allow different circuits to be executed independently of each other. In the NISQ era, this is of particular interest. Due to the hardware limitations, the width and depth of quantum circuits should be reduced as much as possible~\cite{DiAdamo.2022}. In general, validation of use cases respectively the aforementioned computationally intensive queries is only possible with small problem instances in the NISQ era.
\section{Experimentation}\label{sec:trial}

In this Section, we describe the experiment to test the hybrid architecture. For the development of the first prototype, we use \textit{Python}, \textit{FastAPI}, \textit{Docker} and \textit{Qiskit}~\cite{Qiskit} as implementation technologies. The objective is to verify the functioning of the reference procedure described in Section ~\ref{sec:architecture}. To validate the procedure, we resort to well-known quantum routines. We first consider the following Table~\ref{tab:data}.

\begin{table}[!h]
\renewcommand{\arraystretch}{1.3}
\caption{Test Data} \label{tab:data}
\centering
\begin{tabular}[t]{l c S[table-format=-1.2] S[table-format=-1.2] c}
\hline
& {\bfseries ID} & {\bfseries Feature1} & {\bfseries Feature2} & {\bfseries Cluster} \\
\hline
Centroid A&0&-0.5&0.5&blue\\
Centroid B&1&0.2&-0.2&green\\
Data Point &2&0.15&-0.15&\textbf{?}\\
Data Point&3&-0.45&0.45&\textbf{?}\\
\hline
\end{tabular}
\end{table}

This represents some relational data. Each row (tuple) represents a data point that has two properties (called features) \textit{Feature1} and \textit{Feature2}. Each tuple also has a unique ID. The first two tuples are data points, each representing a cluster centroid. The remaining data points are now to be assigned to a cluster. We aim to achieve this with the support of a routine or algorithm for distance estimation as used in the Quantum K-Means algorithm~\cite{Ouedrhiri.2021,DiAdamo.2022}. 

According to the reference procedure in Section~\ref{sec:architecture}, we implement the following services and functionalities in the prototype:

\begin{enumerate}
\item The \textit{Decision Service} receives a request from the \textit{Application} for the assignment of data points to clusters and decides to solve this problem using a \textit{Gate-based Quantum Computer}. The \textit{Application} is informed of this.

\item Next, a defined amount of data records is extracted via the \textit{Data Service}.

\item The \textit{Circuit Service} creates several circuits. We create a circuit for each pair of data point and centroid. In this first experiment, we have opted for this simple approach. First, the data provided is encoded. In our case, these are the features and the ID of the data points. We need the latter for post-processing (cf. step 5). For the encoding of the features we use the so-called \textit{angle embedding}~\cite{DiAdamo.2022,Ouedrhiri.2021}. For this we calculate per tuple the two angles $\theta=(\mathit{Feature1} + 1)\frac{\pi}{2}$ and $\varphi=(\mathit{Feature2} + 1)\frac{\pi}{2}$, whereby angle values between $0$ and $\pi$ are allowed. For the encoding of the IDs we use the basis encoding in this example~\cite{Weigold.2022}. Next the distance estimation algorithm follows, which works on the encoded information (cf. Subsection~\ref{sec:trial_alg}). Note: For the encoding of the IDs, we have chosen basis encoding in this first example. Other implicit encoding types (as in~\cite{Matteo.2020}) are also of interest. In addition, the identification of records in a distributed database system requires more comprehensive identifiers than just an ID as in this example. This was reported in~\cite{Zajac.2022}.

\item The generated circuits are transmitted to a \textit{Gate-based Quantum Computer} via the \textit{Backend Service} and their execution is triggered. In our case, it is the quantum cloud service from IBM Quantum\footnote{IBM Quantum. \url{https://quantum.ibm.com/}, 2023}, which we used for our tests.

\item In the final step, the \textit{Result Manager} checks the execution status and retrieves the result after a circuit has been executed. The result is then forwarded to the \textit{Application}. The corresponding cluster of each data point can be derived from the overall result (comparison of all points with both centroids, cf. Subsection~\ref{sec:trial_results}). The IDs of these data points are included in the individual results. The \textit{Application} is responsible for post-processing the overall result. Based on the IDs, the Table~\ref{tab:data} can be supplemented with the information of the correct cluster. In a relational database, we can imagine that this table is indexed by ID, so that records can be efficiently identified by ID for an update.
\end{enumerate}

\subsection{Circuit}\label{sec:trial_alg}

\begin{figure*}[ht]
 \centering
 \includegraphics[scale=0.75]{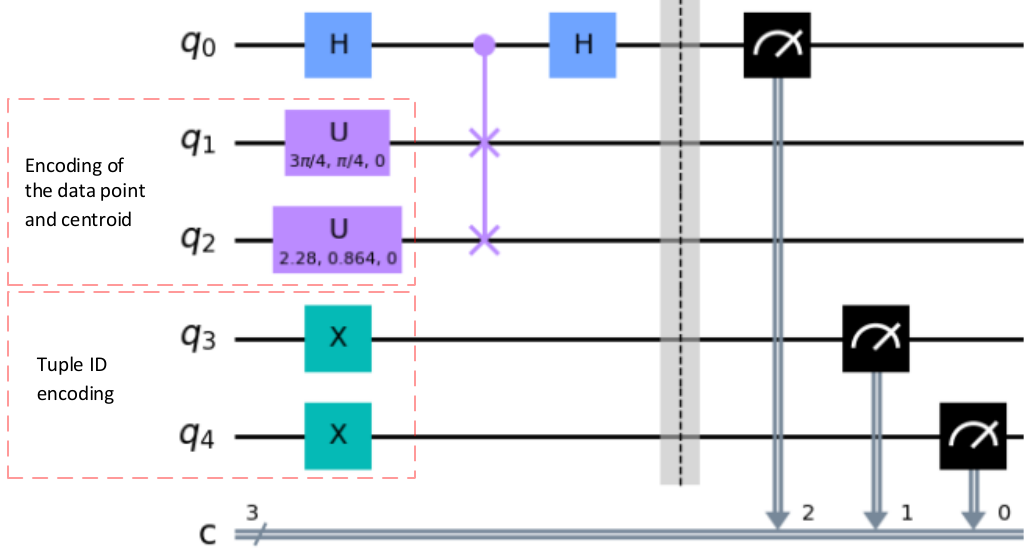}
 \caption{Distance Estimation Algorithm following~\cite{Ouedrhiri.2021} with additional encoding of the Tuple IDs to utilize them in post-processing. Parameterized with Data Point ID=3 and Centroid A. Drawn by \textit{Qiskit}.} \label{fig:alg}
\end{figure*}

The circuit created in Step~3) can be roughly divided into two sections \textit{Encoding} and \textit{Algorithm}. Figure~\ref{fig:alg} shows the circuit as an example for a data point centroid pair. The encoding is marked. To encode the features, so called $U_{3}(\theta,\varphi,0)$ gates can be used. The parameters of the gates are the previously calculated angles. \textit{Pauli X} gates can be used for encoding the IDs. In this case, an ID is represented by a corresponding bit string. 

The core of the actual algorithm is the estimation of the distance between a data point and a centroid. The basis of the algorithm is described in~\cite{Ouedrhiri.2021}. We supplement it with the encoding of the IDs. The distance is reflected in the measurement result and how often a particular measurement result occurs. We will deal with this in Subsection~\ref{sec:trial_results}.

\subsection{Results}\label{sec:trial_results}
In this Subsection we review the results, which are shown in the Table~\ref{tab:res1}. 

\begin{table}[h]
\renewcommand{\arraystretch}{1.4}
\caption{Experiment Results} \label{tab:res1}
\centering
\begin{tabular}[t]{lc|c|c|cc}
\hline
&\bfseries \makecell{Data Point\\ID}
&\bfseries \makecell{Bit sequence\\ for $\mathbf{c[2]=1}$} &\bfseries \makecell{Determined\\ Frequency} &\bfseries \makecell{Calculated\\ Frequency}\\
\hline
Centroid A&2&\textbf{1}10&43&$\approx 47$\\
Centroid B&2&-&0&$\approx 0.513$\\
Centroid A&3&-&0& $\approx 0.061$\\
Centroid B&3&\textbf{1}11&43&$\approx 46$\\
\hline
\end{tabular}
\end{table}

Each data point (\textbf{Column~1}) is compared to the two centroids to determine the proximity in each case. In quantum-based distance estimation, the estimated distance between points can be derived from the frequency of a particular state that we obtain from the measurements. For this purpose, we use the \textit{ibmq\_qasm\_simulator} with $1000$ shots (number of repetitions) per centroid/data point comparison. According to the Figure~\ref{fig:alg}, we measure $q_{0} \rightarrow c[2]$, $q_{3} \rightarrow c[1]$, and $q_{4} \rightarrow c[0]$ and obtain in this way a bit sequence representing the measured state (\textbf{Column~2}). $c[1]c[0]$ corresponds to the binary representation of the ID of the data point. We are interested in the case $c[2]=1$ (the particular state) and the frequency of its occurrence (\textbf{Column~3}). In this case, a low frequency correlates with a low estimated distance between a point and a centroid (if the value for the determined frequency is zero, then it means that the probability of measuring the particular state is low). \textbf{Column~4} contains the calculated frequencies (for $1000$ repetitions). We need these to compare the determined values (in Column~3) with predicted values and thus verify the correctness of the architectural workflow. The calculated probability for $c[2]=1$ is: $\frac{1}{2}-\frac{1}{2}|\braket{q_{1}|q_{2}}|^2$~\cite{Ouedrhiri.2021,DiAdamo.2022}. The calculated frequencies correspond approximately to the determined values and prove the correct functioning of the architecture implementation.

Based on the estimated distances, the source table can be supplemented. The supplemented entries are shown in Table~\ref{tab:res3}.

\begin{table}[h]
\renewcommand{\arraystretch}{1.3}
\caption{Updated Test Data} \label{tab:res3}
\centering
\begin{tabular}[t]{l c S[table-format=-1.2] S[table-format=-1.2] c}
\hline
& {\bfseries ID} & {\bfseries Feature1} \ & {\bfseries Feature2} & {\bfseries Cluster} \\
\hline
\dots&\dots&\dots&\dots&\dots\\
Data Point&2&0.15&-0.15&\textbf{green}\\
Data Point&3&-0.45&0.45&\textbf{blue}\\
\hline
\end{tabular}
\end{table}

The architecture thus allows data from a database to be processed by a quantum algorithm to assign data points to the correct clusters. The correct generation of the corresponding circuits was ensured by the previously mentioned comparison of determined and calculated frequencies.
\section{Summary and Future Work}\label{sec:summary}

In this paper, we have presented a Hybrid Data Management Architecture to enable data exchange between databases and quantum computers. We have experimentally proven a correct functioning of the architecture. In the experiments, quantum circuits were created and executed based on relational data. The circuits implement an algorithm for the distance estimation between data point and centroid and includes the encoding of the classical data. However, the architecture is not limited to this use case.

Next, we would like to leverage the architecture to implement more data-centric applications. In doing so, we would like to encode data organized in hierarchical structures (for example, labelled trees for which we have outlined an initial idea~\cite{DBLP:conf/vldb/Zajac23}, or graphs) in order to execute selected queries on the encoded data using appropriate quantum algorithms. The implemented routines are also accompanied by time and space complexity analysis to deduce any advantages over purely classical solutions.

\section*{Acknowledgement}
This work has been funded by Deutsche Forschungsgemeinschaft (DFG, German Research Foundation) grant \#385808805. We would like to thank Stefanie Scherzinger from University of Passau for many prolific discussions as well as helpful suggestions.

\bibliographystyle{plainnat}
\bibliography{references}

\end{document}